\documentclass[10pt,twocolumn,letterpaper]{article}

\usepackage{cvpr}              

\usepackage{graphicx}
\usepackage{amsmath}
\usepackage{amssymb}
\usepackage{booktabs}
\usepackage{multirow}
\usepackage{comment}
\usepackage[pagebackref,breaklinks,colorlinks]{hyperref}

\usepackage[capitalize]{cleveref}
\crefname{section}{Sec.}{Secs.}
\Crefname{section}{Section}{Sections}
\Crefname{table}{Table}{Tables}
\crefname{table}{Tab.}{Tabs.}

\begin{document}

\title{AI-Synthesized Voice Detection Using Neural Vocoder Artifacts}

\author{ Chengzhe Sun \hspace{0.7cm} Shan Jia \hspace{0.7cm} Shuwei Hou\hspace{0.7cm} Siwei Lyu\\
University at Buffalo, State University of New York, NY, USA\\
{\tt\small \{csun22, shanjia, shuweiho, siweilyu\}@buffalo.edu}
}

\maketitle

\begin{abstract}
Advancements in AI-synthesized human voices have created a growing threat of impersonation and disinformation, making it crucial to develop methods to detect synthetic human voices. This study proposes a new approach to identifying synthetic human voices by detecting artifacts of vocoders in audio signals. Most DeepFake audio synthesis models use a neural vocoder, a neural network that generates waveforms from temporal-frequency representations like mel-spectrograms. By identifying neural vocoder processing in audio, we can determine if a sample is synthesized. To detect synthetic human voices, we introduce a multi-task learning framework for a binary-class RawNet2 model that shares the feature extractor with a vocoder identification module. By treating vocoder identification as a pretext task, we constrain the feature extractor to focus on vocoder artifacts and provide discriminative features for the final binary classifier. Our experiments show that the improved RawNet2 model based on vocoder identification achieves high classification performance on the binary task overall. Codes and data can be found at \url{https://github.com/csun22/Synthetic-Voice-Detection-Vocoder-Artifacts}.
\end{abstract}

\section{Introduction}
\label{sec:intro}

In recent years, the rapid development of AI technologies, particularly deep learning, has resulted in a surge of synthetic media, commonly referred to as "DeepFakes." These media are highly realistic and can be challenging to distinguish from genuine content, making them a significant concern. While AI-synthesized still images and videos have received much attention, synthetic human voices have also undergone significant advances, achieving unprecedented quality and efficiency. These voices have the potential to revolutionize voice-based user interfaces for intelligent home assistants and wearable devices and can even help individuals who have lost their ability to speak due to conditions like strokes or Amyotrophic Lateral Sclerosis (ALS).

However, the increasing realism and availability of synthetic human voices also pose significant risks. Scammers have used AI-synthesized voices to impersonate individuals and deceive others into transferring money or providing sensitive information. In one instance, a scammer used an AI-synthesized voice to impersonate a UK company's CEO and tricked an employee into transferring a large sum of money to the scammer's account \cite{forbes}. Moreover, trolls on the internet have used free AI voice cloning tools to imitate the voices of celebrities and create content ranging from memes to virulent hate speech \cite{elevenlabs}.

While methods to detect AI-synthesized images and videos have been extensively studied, methods to detect synthetic human voices have received less attention and are underdeveloped. This is because audio signals have different characteristics that make it difficult to apply image-based detection methods. Early detection methods often analyze statistical features unique to audio signals. For example, \cite{albadawy2019detecting} compares higher-order statistics in the bi-spectral domain that capture local phase inconsistencies in synthetic voices. Recent studies \cite{tak2021end, lv2022fake, xue2023learning} tend to use well-designed models for automatic and comprehensive feature learning to detect synthesized audio.

This work proposes a new approach to detecting synthetic human voices based on artifacts introduced by the neural vocoders used in the generation process. A neural vocoder is a specialized neural network that synthesizes audio waveforms from temporal-frequency representations like mel-spectrograms. Since neural vocoders are the final step in most AI-based audio synthesis models, it is unlikely that real audio signals will be processed with neural vocoders. Thus, the vocoder artifacts can provide cues to identify synthetic human voices.

\begin{figure*}[t]
    \centering
    \includegraphics[width=1.0\textwidth] {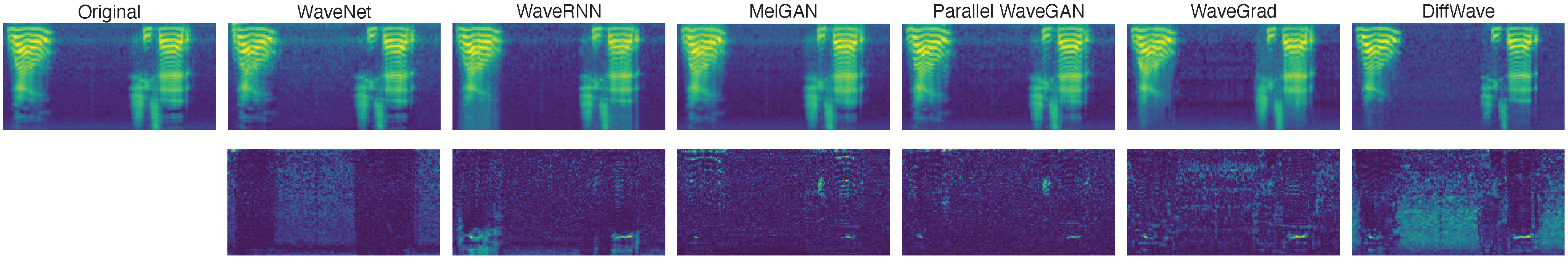}
    \vspace{-0.7cm}
    \caption{\small The artifacts introduced by the neural vocoders to a voice signal. We show the mel-spectrogram of the original (top left) and the self-vocoded voice signal (top right five). Their differences corresponding to the artifacts introduced by the vocoder are shown at the bottom.} 
    \vspace{-0.5cm}
    \label{fig:sig_diff}
\end{figure*}
We aim to highlight the distinct signal artifacts left by neural vocoders in synthetic audio signals. The foremost objective of this study is to explore the artifacts of vocoders. To do this, we constructed a dataset called {\em LibriSeVoc}, which controls for other factors and only probes for the vocoder signature. The dataset contains evenly distributed data for various vocoders. We used six different neural vocoders to create the LibriSeVoc dataset to reflect the diversity in architecture and mechanisms of neural vocoders. "Self-vocoding" samples were sourced from the same original audio signals to highlight the artifacts introduced by the vocoders. Figure \ref{fig:sig_diff} shows the differences in the mel-spectrogram of one original voice and its self-vocoded voice signals. Visible artifacts introduced by different neural vocoder models can be observed, which serve as the basis of our detection algorithm. While these artifacts may be subtle to visualize, this work demonstrates that they can be captured by a trained classifier.

To take advantage of the vocoder artifacts in detecting synthetic human voices, we developed a multi-task learning strategy. We used a binary classifier that shares the front-end feature extractor (e.g., RawNet2 \cite{tak2021end}) with the vocoder identification module. This accommodates the insufficient number of existing real and synthetic human voice samples by including the self-vocoding samples in LibriSeVoc as additional training data. We treated vocoder identification as a pretext task to constrain the front-end feature extraction module to focus on vocoder-level artifacts and build highly discriminative features for the final binary classifier. Our experiments showed that our RawNet2 model achieved outstanding classification performance on our LibriSeVoc dataset and two public DeepFake audio datasets. We also evaluated our method under different post-processing scenarios and demonstrated good detection robustness to re-sampling and background noise.

The main contributions of our work are as follows:

\begin{itemize}
\item We propose to focus on neural vocoder artifacts as specific and interpretable features for detecting AI-synthesized audio;
\item We designed a novel multi-task learning approach that combines a binary classification task with a vocoder identification module. This approach constrains the feature extractor to learn discriminative vocoder artifacts for detecting synthetic human voices;
\item We provide LibriSeVoc as a dataset with self-vocoding samples created using six state-of-the-art vocoders to highlight and exploit the vocoder artifacts;
\item Our proposed method was experimentally evaluated on three datasets and demonstrated its effectiveness.
\end{itemize}

Overall, our work provides a new and promising approach to detecting synthetic human voices by focusing on neural vocoder artifacts. Our multi-task learning strategy, together with the LibriSeVoc dataset, could serve as a valuable resource for future research in this area.

\section{Related Works}
\vspace{-0.1cm}
In this section, we provide a literature review that is relevant to our research, including voice synthesis methods, state-of-the-art neural vocoder models, and existing AI-synthesized voice detection methods. 

\subsection{Human Voice Synthesis} 

The synthesis of human voice is a significant challenge in the field of artificial intelligence, with various practical applications such as voice-driven smart assistants and accessible user interfaces. Human voice synthesis can be classified into two general categories: text-to-speech (TTS) and voice conversion (VC). In this work, we focus on recent TTS and VC methods that use deep neural network models.

TTS systems transform input text into audio using the target voice and typically consist of three components: a text analysis module that converts the input text into linguistic features, an acoustic model that generates acoustic features in the form of a mel-spectrogram from the linguistic features, and a vocoder. Recent TTS models based on deep neural networks include WaveNet~\cite{vanwavenet}, Tacotron~\cite{DBLP:journals/corr/WangSSWWJYXCBLA17}, Tacotron 2~\cite{9555268}, ClariNet~\cite{ping2018clarinet}, and FastSpeech 2s~\cite{ren2020fastspeech}.

In contrast, VC models take a sample of one subject's voice as input and create output audio of another subject's voice of the same utterance. Recent VC models (\eg,~\cite{doi:10.1002/mp.12752,chen2014voice,mohammadi2014voice}) usually work within the mel-spectrum domain and use deep neural network models to map between the mel-spectrograms of the input and output voice signals. These models use neural style transfer methods such as variational auto-encoder (VAE) or generative adversarial network (GAN) models to capture the utterance elements in the input voice and then combine them with the style of the output voice. The resulting mel-spectrogram is then reconstructed to an audio waveform using a neural vocoder. Both the TTS and VC models employ deep neural network models trained on large-scale human voice corpora.

\subsection{Neural Vocoders}
\label{subsec:voc}

Vocoders are crucial components in both TTS and VC models as they synthesize output audio waveforms from mel-spectrograms. However, the transformation from audio waveforms to mel-spectrograms leads to the loss of information due to binning and filtering, making it difficult to recover the audio waveform from a mel-spectrogram. In recent years, deep neural network-based vocoders have been developed, significantly improving training efficiency and synthesis quality. There are three main categories of existing neural vocoders: autoregressive models, diffusion models, and GAN-based models.

Autoregressive models are probabilistic models that predict the distribution of each audio waveform sample based on all previous samples. However, since this process involves linear sample-by-sample generation, autoregressive models are slower than other methods. WaveNet~\cite{vanwavenet}, the first autoregressive neural vocoder, can also serve as a TTS or VC model depending on the input. WaveRNN~\cite{kalchbrenner2018efficient} is another autoregressive vocoder that uses a single-layer recurrent neural network to efficiently predict 16-bit raw audio samples from mel-spectrogram slices.

Diffusion models are probabilistic generative models that run diffusion and reverse processes. The diffusion process is characterized by a Markov chain, which gradually adds Gaussian noise to an original signal until the noise is eliminated. The reverse process is a de-noising stage that removes the added Gaussian noise and converts a sample back to the original signal. WaveGrad~\cite{chen2020wavegrad} and DiffWave~\cite{kong2020diffwave} are two notable examples of diffusion-based vocoder models. While diffusion models are the most time-efficient vocoders, their reconstruction qualities are inferior to autoregressive models, and the generated samples may contain higher levels of noise and artifacts.

GAN-based models follow the generative adversarial network (GAN) architecture~\cite{goodfellow2014generative}, which employs a deep neural network generator to model the waveform signal in the time domain and a discriminator to estimate the quality of the generated speech. Mel-GAN~\cite{kumar2019melgan} and Parallel WaveGAN~\cite{yamamoto2020parallel} are the two most commonly used GAN-based neural vocoders. Recent works have shown that GAN-based vocoders outperform autoregressive and diffusion models in both generation speed and generation quality.

\subsection{AI-synthetic Human Voice Detection}

In recent years, detecting synthetic human voices has become crucial due to their potential misuse. While extensive research has focused on audio authentication for speech synthesis and replay attack detection~\cite{wu2015spoofing, patil2018survey}, detecting AI-generated audio with high realism and varying models is a developing field. One of the earliest methods for detecting AI-synthetic audio is bi-spectral analysis~\cite{albadawy2019detecting}. This method captures subtle inconsistencies in local phases of synthetic human voices. Real human voice signals have random local phases due to audio waves transmitting and bouncing around in the physical environment, while synthetic human voices do not have these characteristics. Although these local phase inconsistencies cannot be detected by the human auditory system, they can be identified through bi-spectral analysis. Another method, known as DeepSonar~\cite{wang2020deepsonar}, uses network responses of audio signals as the feature to detect synthetic audio. The ASVspoof Challenge 2021 evaluates additional state-of-the-art synthetic voice detection methods. The Gaussian mixture models CQCC-GMM~\cite{todisco2019asvspoof}, LFCC-GMM~\cite{todisco2019asvspoof}, a light convolutional neural network model LFCC-LCNN~\cite{todisco2019asvspoof}, and RawNet2~\cite{tak2021end} have achieved the most reliable performance as primary baseline algorithms.

\begin{figure*}[t]
  \centering 
  \vspace{-0.2cm}
\includegraphics[width=\linewidth]{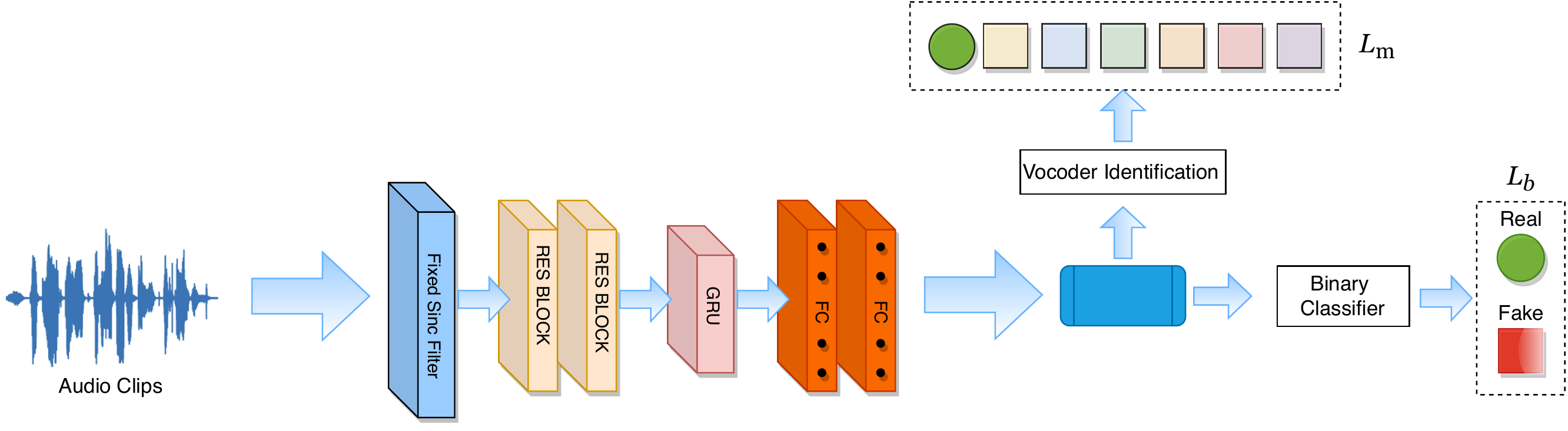}
  \caption{Framework of the proposed synthesized voice detection method. Rather than simply learning discriminative features for binary classification, we incorporate a vocoder identification module with a multi-class classification loss to direct the feature extraction network toward prioritizing vocoder artifacts.} 
  \vspace{-0.3cm}
  \label{fig:method}
 \end{figure*}

Recent studies have focused on improving the generalization capacity of fake audio detection. Various well-designed models have been proposed for Deepfake audio detection, such as the spectro-temporal graph attention network~\cite{tak2021end}, unsupervised pretraining models~\cite{lv2022fake}, biometric characteristics verification model~\cite{pianese2022deepfake}, and self-distillation framework~\cite{xue2023learning}. However, Müller \etal~\cite{muller2022does} evaluated twelve architectures on their dataset with $37.9$ hours of audio recording and found that related work performs poorly on real-world data, with some models even degenerating to random guessing. Thus, there is a high demand for developing efficient and effective models for AI-synthesized audio detection.

\section{Method}
\vspace{-0.2cm}
\label{sec:RawNet2}

We aim to detect synthetic human voices by identifying vocoder artifacts present in the audio signals. Since real human voice signals typically do not have vocoder artifacts, except for our self-vocoding signals that are specifically designed to have them, identifying the presence of vocoder artifacts is a key feature in detecting synthetic human voices.

To achieve this, let $\mathbf{x}$ be the waveform of a human voice signal with a label $y \in {0, 1}$, where $0$ corresponds to a real human voice and $1$ corresponds to a synthetic human voice. Our goal is to build a classifier $\hat{y} = F_\theta(\mathbf{x})$ that predicts the label of an input $\mathbf{x}$. We utilize the recent RawNet2 model \cite{tak2021end} as the backbone for our classifier, as it was designed to operate directly on raw waveforms. This reduces the risk of losing information related to neural vocoder artifacts when compared to using pre-processed features such as mel-spectrograms or linear frequency cepstral coefficients (LFCCs).

The binary detection model can be constructed as a cascade of neural networks 
\begin{equation}
    F_\theta(\mathbf{x}) = B_{\theta_B}(R_{\theta_R}(\mathbf{x}))
\end{equation}
where $R_{\theta_R}(\mathbf{x})$ is the front-end RawNet2
model for feature extraction with its own set of parameters $\theta_R$, $B_{\theta_B}$
is a back-end binary classifier and $\theta_B$ are its specific parameters, with
$\theta = (\theta_R, \theta_B)$. We can train this classifier directly, as in the
previous work \cite{yang2021multi}, by solving 
\begin{equation}
    \min_\theta \sum_{(\mathbf{x},y) \in T} L_\text{b}(y, F_\theta(\mathbf{x}))
\end{equation}
where $L_\text{b}(y,\hat{y})$ could be any loss function for binary classification, for instance, the cross-entropy loss. The variable $T$ refers to the training dataset that contains labeled real and synthetic human voice samples. However, this method assumes that there is a large number of synthetic human voice samples available, which is increasingly difficult to achieve due to the rapid advancement of synthesis technology. Additionally, this approach does not take into account the unique statistical properties of neural vocoders, which can be an essential indicator for synthetic audio signals.

To address the aforementioned problem, we propose a multi-task learning approach that combines binary classification with a vocoder identification task. This approach is designed to emphasize the importance of identifying vocoder-level artifacts in synthetic audio signals. Specifically, we augment our detection model with a vocoder identifier $M_{\theta_M}$, which categorizes a synthetic voice into one of the $c$ possible neural vocoder models ($c\in[0, C]$ where $C\ge 2$). Our goal is to ensure that the feature extractor is trained to capture the distinct statistical characteristics of vocoders, making it more sensitive to these features. This approach is similar to self-supervised representation learning \cite{SSRL}. To this end, we form a new classification objective, as
\begin{equation}
    \begin{array}{l}
\min_{\theta_B,\theta_R} \lambda\sum_{(\mathbf{x},y) \in T} L_\text{b}(y, B_{\theta_B}(R_{\theta_R}(\mathbf{x}))) \\
+ \min_{\theta_M,\theta_R} (1-\lambda)\sum_{(\mathbf{x},c) \in T'} L_\text{m}(c, M_{\theta_M}(R_{\theta_R}(\mathbf{x})))
\end{array}
\end{equation}
In this equation, $L_\text{m}$ is a multi-class loss function, and we use the softmax loss in our experiments. $T'$ is a dataset containing synthetic human voices created with different neural vocoders as corresponding labels. This dataset is much easier to create by performing ``self-vocoding'', \ie, creating synthetic human voices by running real samples through the mel-spectrogram transform and inverse, the latter performed with neural vocoders. We created such a dataset, LibriSeVoc, which will be described in detail in Section \ref{sec:data-set}. $\lambda$ is an adjustable hyper-parameter that controls the trade-off between the two loss terms. 

The whole framework of our detection model is shown in Figure \ref{fig:method}. Note that the two classification modules in the new learning objective function serve different roles. The first term pertains to binary classification and aims to distinguish between authentic and fake audio. Meanwhile, the second term focuses on vocoder identification, serving as a pretext task to direct the feature extractor's attention toward vocoder-related artifacts. 
The two tasks share the feature extraction component so that the distinct features of the vocoders can be captured and transferred to the binary classification task. 


\begin{table*}[!t]
\centering
\footnotesize
\caption{Data details of voice samples from real and each vocoder category in the LibriSeVoc dataset.}
\label{tab:LibriSeVoc-stat}
\setlength{\tabcolsep}{6pt}
\begin{tabular}{l|c|c|c|c|c|c|c|c} 
\hline
\textbf{Model}          &\textbf{Train-hour} &\textbf{Train-sample} &\textbf{Dev-hour}      &\textbf{Dev-sample} &\textbf{Test-hour} &\textbf{Test-sample}   &\textbf{Total-hour} &\textbf{Total-smaple} \\
\hline
Real                     &20.95 &7,920 & 6.97 &2,640 &7.00 &2,641 &34.92 & 13,201\\
\hline
WaveNet (A01)            & 20.87 &7,920 & 6.97 &2,640 &6.91  &2,641  &34.77   &13,201\\
WaveRNN (A02)            & 20.95 &7,920 & 7.00 &2,640 &6.94  &2,641  &34.90  &13,201 \\
WaveGrad (D01)           & 20.98 &7,920 & 7.01 &2,640 &6.95  &2,641  &34.95   &13,201 \\
DiffWave (D02)           & 20.98 &7,920 & 7.01 &2,640 &6.95  &2,641  &34.94   &13,201 \\
MelGAN (G01)             & 20.76 &7,920 & 6.94 &2,640 &6.88  &2,641  &34.59   &13,201 \\
Parallel WaveGAN (G02)   & 20.76 &7,920 & 6.94 &2,640 &6.88  &2,641  &34.59   &13,201 \\

\hline
\textbf{Total}           &146.25 &55,440 &48.84 &18,480 &48.51 &18,487   &243.66 &92,407 \\
\hline
\vspace{-0.4cm}
\end{tabular}
\end{table*}

\section{Experiments}

In this section, we present a series of experiments to evaluate the effectiveness of our proposed synthetic voice detection method. We start by introducing our LibriSevoc dataset, as well as two publicly available datasets that we use for evaluation. Then, we compare our method with state-of-the-art models on all three datasets, using both intra- and cross-dataset testing scenarios. Finally, we examine the robustness of our detection models to common post-processing operations.

\subsection{Datasets}
\label{sec:data-set}
Three DeepFake audio datasets are considered in experiments, namely our LibriSeVoc, and two public datasets WaveFake~\cite{frank2021wavefake} and ASVspoof 2019~\cite{todisco2019asvspoof}. 



\noindent{\bf LibriSeVoc Dataset.} 
We have created a new open-source dataset, named LibriSeVoc, for the task of vocoder artifact detection. The statistical features of neural vocoders have not been extensively studied before, and the availability of large-scale datasets for the task of vocoder identification, particularly those that include multiple types of vocoders, is limited. 
We derived the LibriSeVoc dataset from the widely used LibriTTS speech corpus \cite{zen2019libritts}, which is often utilized in text-to-speech research \cite{kim2020glow,valle2020flowtron,chen2020multispeech}. The LibriTTS corpus is based on the Librispeech dataset \cite{panayotov2015librispeech}, which contains samples extracted from audiobooks available on LibriVox~\cite{librivox}.

\begin{table*}[t]
\centering
\footnotesize
\caption{Details of evaluation datasets.}
\vspace{-0.2cm}
\setlength{\tabcolsep}{10pt}
\begin{tabular}{l|c|c|c|c|c} 
\hline
\textbf{Dataset}    & \textbf{\#Vocoder type}  & \textbf{Frequency} &\textbf{Training size }  & \textbf{Dev size }  &  \textbf{Testing size} \\
\hline
LibriSeVoc    &6  &24kHz  &55,440  &18,480 &18,487 \\ 
WaveFake Dataset~\cite{frank2021wavefake} &6  &16kHz  &64,000 &16,000 &24,800 \\
ASVspoof 2019~\cite{lavrentyeva2019stc}  &6  &16kHz  &25,380  &24,844 &71,237 \\
\hline
\end{tabular}
\label{tab:db}
\end{table*}

We consider six state-of-the-art neural vocoders to generate speech samples in the LibriSeVoc dataset, namely, WaveNet and WaveRNN from the autoregressive vocoders, Mel-GAN and Parallel WaveGAN from the GAN-based vocoders, and WaveGrad and DiffWave from the diffusion-based vocoders. Specifically, we have 34.92 hours of real audio samples and have generated self-vocoded audio using six vocoders for each sample. A total of $208.74$ hours of synthesized samples are created in the dataset. 

Specifically, each vocoder synthesizes waveform samples from a given mel-spectrogram extracted from an original sample; we refer to this process as ``self-vocoding.'' By providing each vocoder with the same mel-spectrogram, we ensure that any unique artifacts present in the synthesized samples are attributable to the specific vocoder used to reconstruct the audio signal. 
The statistical information of the LibriSeVoc dataset is summarized below.

\begin{figure}[htbp]
  \centering 
  \vspace{-0.5cm}
\includegraphics[width=\linewidth]{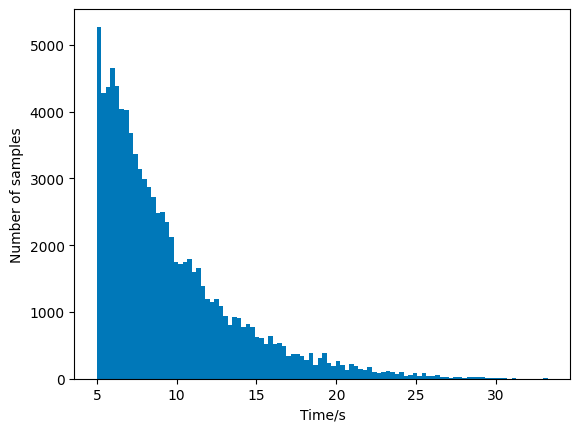}
  \caption{Histogram of audio samples length in LibriSeVoc.}  \vspace{-0.2cm}
  \label{fig:length}
 \end{figure}

\begin{itemize} 
\item The dataset contains $13,201$ real audio samples and $79,206$ fake audio samples from six vocoders (each with $13,201$ audio samples). 
\item A total of $92,407$ audio samples are included in the dataset, with audio length from 5 seconds to 34 seconds at 24kHz. Figure \ref{fig:length} shows the detailed audio length distribution in the dataset.
\item We further split the whole dataset into three non-overlapped sets for training ($55,440$ samples), development ($18,480$ samples), and testing ($18,487$ samples) at the ratio of 6:2:2. More details can be found in Table \ref{tab:LibriSeVoc-stat}.
\end{itemize}

\noindent{\bf WaveFake Dataset.} This dataset~\cite{frank2021wavefake} collects DeepFake audios from six vocoder architectures, including MelGAN, FullBand-MelGAN, MultiBand-MelGAN, HiFi-GAN, Parallel WaveGAN, and WaveGlow. It consists of approximately 196 hours of generated audio files derived from the LJSPEECH~\cite{ito2017LJspeech} dataset. Note that five of the six vocoder models in WaveFake are GAN-based vocoders. Differently, our LibriSeVoc dataset aims to consider a high diversity of vocoders for artifact extraction and covers three categories of widely-used vocoder structures, including autoregressive models, diffusion models, and GAN-based models.

\noindent{\bf ASVspoof 2019 Dataset.} This dataset~\cite{lavrentyeva2019stc} is derived from the VCTK base corpus~\cite{veaux2016superseded}, which includes speech data captured from 107 speakers. It contains three major forms of spoofing attacks, namely synthetic, converted, and replayed speech. We labeled the samples from different vocoders as different classes in the training set for the multi-class loss calculation. More details about the dataset can be found in Table \ref{tab:db}.


\subsection{Implementation Details}
We utilize the RawNet2~\cite{tak2021end} model as the backbone for feature learning. The Adaptive Moment Estimation (Adam)~\cite{kingma2014adam} is used as our optimizer with a learning rate of 0.0001 and a batch size of 32.  
The loss weight $\lambda$ is set as 0.5 in the experiment. To report the detection performance, we calculate the Equal Error Rate (EER) following previous studies~\cite{frank2021wavefake, todisco2019asvspoof, yamagishi2021asvspoof}.

\subsection{Baseline methods}
\noindent{\bf LFCC-LCNN~\cite{lavrentyeva2019stc}.} 
The LFCC-LCNN method combines LFCC feature extraction with an LCNN classifier (DNN). LFCC-LCNN has been widely used since ASVspoof 2019. It achieved the second-best performance in the ASVspoof 2021 Speech Deepfake track during the ASVspoof challenge.

\noindent{\bf RawNet2~\cite{tak2021end}.}
The RawNet2 model is based on DNN speaker embedding extraction with the raw waveform as inputs. This powerful model uses a technique named feature map scaling which scales feature maps similar to squeeze-excitation. It performed the best in the ASVspoof 2021 Speech Deepfake track.

\noindent{\bf WavLM~\cite{article}.}
The WavLM model from Microsoft is a self-supervised pre-trained multilingual model that can be used for a variety of downstream speech tasks. It is specifically designed to maintain speech content modeling based on masked speech prediction while also improving the potential for non-ASR tasks through speech denoising. Additionally, WavLM uses a gated relative position bias in its Transformer structure to better capture the sequence ordering of input speech.

\noindent{\bf Wav2Vec2-XLS-R~\cite{unknown}.}
The Wav2Vec2-XLS-R model is a large-scale multilingual pre-trained model for speech-related tasks developed by Meta AI. It uses the wav2vec 2.0 objective for speech representation learning in 128 languages. This model can be fine-tuned for downstream tasks such as automatic speech recognition, translation, or speech classification.
\subsection{Intra-dataset Evaluation}
\noindent{\bf Synthetic Human Voice Detection.} We first report the performance on the main task, \ie, the classification of real and synthetic human voices on the LibriSeVoc and WaveFake datasets. The results in the
first three columns of Table~\ref{tab:com1} show that our multi-task learning-based RawNet2 model with vocoder identification as a pretext task achieves the lowest EER of 0.13\% on LibriSeVoc and 0.19\% on WaveFake, obviously outperforming other baselines re-trained on each dataset. 

\begin{table}[ht]
\centering
\footnotesize
\caption{Detection EER (\%) of intra-dataset testing on three datasets.}
\vspace{-0.2cm}
\setlength{\tabcolsep}{6pt}
\begin{tabular}{l|c|c|c} 
\hline
\textbf{Methods}             & \textbf{LibriSeVoc } & \textbf{WaveFake }  & 
\textbf{ASVspoof} \\
\hline
LFCC-LCNN~\cite{lavrentyeva2019stc}     &0.14           &\textbf{0.19}    &11.60     \\ 
RawNet2~\cite{tak2021end}               &0.17           &0.32     &6.10        \\ 
WavLM~\cite{article}                    &0.45           &2.92     &6.94     \\
Wav2Vec2-XLS-R~\cite{unknown}      &1.54           &2.33     &13.48     \\
\textbf{Ours}                           &\textbf{0.13}  &\textbf{0.19} &\textbf{4.54}     \\
\hline
\end{tabular}
\label{tab:com1}
\end{table}

\noindent{\bf Evaluation on ASVspoof 2019.} We then evaluate our method on the ASVspoof 2019 dataset with spoofed and fake audios. The results in the last column of Table~\ref{tab:com1} show that our model performed the best in detecting various audio spoofing attacks.

\subsection{Cross-dataset Evaluation}
In cross-domain testing, we evaluated the performance of pre-trained detection models trained on our LibriSeVoc dataset on the WaveFake dataset, which contained voice samples not included in the training set. Table~\ref{tab:com4} displays the comparison results with EER for each vocoder of the WaveFake dataset. It is evident that all the methods showed a noticeable degradation in performance during cross-domain testing, indicating poor generalization ability. Our proposed method achieved the lowest EERs in detecting unseen voice samples generated with the same vocoders as those in our LibriSeVoc dataset, specifically MelGAN and Parallel WaveGAN models. However, it is not surprising that the model exhibited poorer performance on other unseen vocoders. This highlights a limitation of our method, particularly when compared to the original Rawnet2, as the vocoder artifacts are not generalized and are constrained by the training vocoder categories.

\begin{table*}[ht]
\centering
\footnotesize
\caption{Detection EER (\%) of cross-dataset testing on WaveFake dataset.}
\vspace{-0.2cm}
\setlength{\tabcolsep}{7pt}
\begin{tabular}{l|c|c|c|c|c|c|c} 
\hline
\multirow{2}{*}{\textbf{Methods}}   &  \multirow{2}{*}{Overall} & \multicolumn{2}{c|}{Seen vocoder}  & \multicolumn{4}{c}{Unseen vocoder} \\
\cline{3-8}
& & MelGAN & Parallel WaveGAN & WaveGlow  &Multi\-Band MelGAN & Full\-Band MelGAN & HiFi\-GAN \\
\hline
LFCC-LCNN~\cite{lavrentyeva2019stc}  &77.47  &30.06 &90.98 &98.40 &97.17 &99.45  &98.02       \\ 
RawNet2~\cite{tak2021end}        &\textbf{25.98} &3.39&{44.31} &\textbf{0.58} &\textbf{23.34} & \textbf{43.82}  &\textbf{33.38}          \\ 
WavLM~\cite{article}              &50 &50 &50 &50 &50 &50 &50        \\
Wav2Vec2-XLS-R~\cite{unknown}     &50 &50 &50 &50 &50 &50 &50        \\
\textbf{Ours}       & 26.95 & \textbf{2.16} &\textbf{35.53} &4.60 & 31.38 & 45.35 &35.80   \\
\hline
\end{tabular}
\label{tab:com4}
\end{table*}

\subsection{Robustness Evaluation} We evaluate the robustness of our detection method against common post-processing operations by constructing an augmented, degraded dataset from the LibriSeVoc test dataset. First, we resample the input speech to intermediate sampling rates (8kHz, 16kHz, 22.05kHz, 32kHz, and 44.1kHz) and then resample back to the original sampling rate (24 kHz). Additionally, we introduce background noise by adding a single pre-recorded crowd noise sample at three SNR values (8dB, 10dB, and 20dB). We randomly choose between the original, resampled, or noisy speech segments, with probabilities of 40\%, 40\%, and 20\%, respectively. We use this augmented dataset to evaluate the robustness of our detection method against these post-processing operations.
 
 \begin{figure}[t]
  \centering   \includegraphics[width=0.7\linewidth]{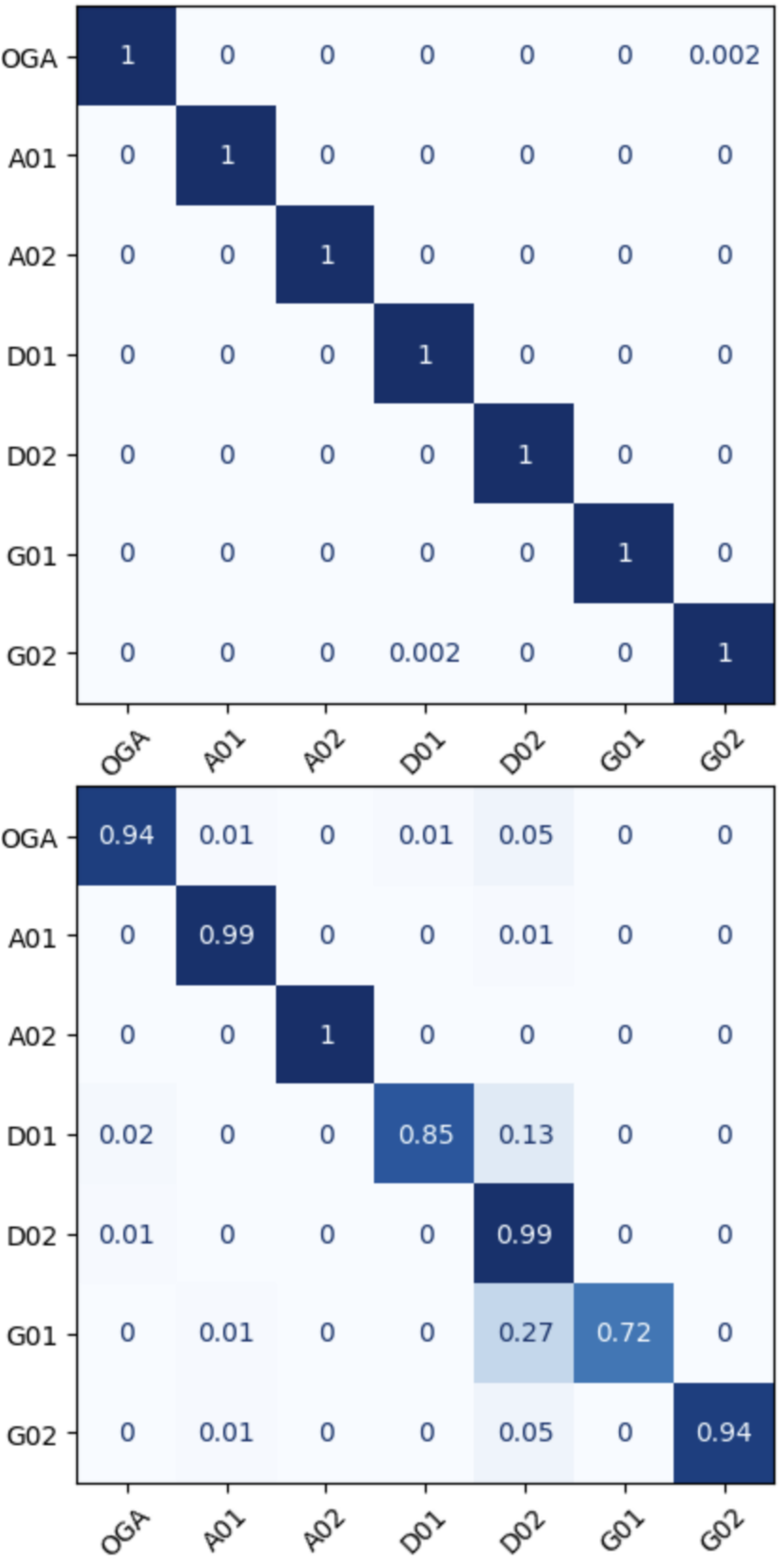}
    \vspace{-0.1cm}
  \caption{Confusion matrices evaluated on LibriSeVoc. Top: original testing set. Bottom: post-processed testing set. (OGA: ground truth, A01: WaveNet, A02: WaveRNN, D01: WaveGrad, D02: Diffwave,   G01: MelGAN, G02: Parallel WaveGAN.) }
  \label{fig: Confusion}
  \vspace{-0.4cm}
 \end{figure} 

The confusion matrix in Figure ~\ref{fig: Confusion} further compares the detection and vocoder identification performance on the original LibriSeVoc set (with a detection EER of 0.13\%) and on the post-processed dataset (with an EER of 2.73\%). 
This shows that our method can extract discriminative vocoder-level features for vocoder identification and is also robust to common data post-processing operations.

\section{Conclusion}

In this work, we propose a novel approach for detecting synthetic human voices by identifying the traces of {\em neural vocoders} in audio signals. To leverage the vocoder artifacts for synthetic human voice detection, we introduce a binary-class RawNet2 model that shares the front-end feature extractor with the one for vocoder identification. We employ a multi-task learning strategy where vocoder identification serves as a pretext task to constrain the front-end feature extraction module for building the final binary classifier. Our experiments demonstrate that our method achieves a high classification performance overall.

While our proposed approach has shown promising results, there is still room for improvement and several areas for future work. Firstly, we plan to expand the LibriSeVoc dataset to include a wider variety of real audio signals and environments. This will help to increase the diversity of the dataset and improve the generalization capability of our model. Secondly, although our method is effective in identifying synthetic audio through vocoder artifacts, we acknowledge that it is an indirect approach. In future work, we aim to explore methods that can directly differentiate real and synthetic audio by combining cues from vocoders and other signal features of audio DeepFakes. Finally, we will explore the use of our approach for detecting and preventing the misuse of synthetic audio in various applications, such as voice cloning or DeepFake videos.

\section*{Social Impact Statement}

The quality and efficiency of synthetic human voices generated by AI models have reached unparalleled heights. However, with the increasing realism and accessibility of these voices, there are significant risks involved. Fraudsters have exploited AI-generated voices to impersonate individuals and trick others into sharing sensitive information or transferring money. In this work, we propose a method that detects vocoder artifacts, which can expose AI-synthesized voices and mitigate the risks associated with them.

To ensure transparency, codes and data related to this work are made available as Open Source at \url{https://github.com/csun22/Synthetic-Voice-Detection-Vocoder-Artifacts}.
\paragraph{Acknowledgment.} This work is supported by the Center for Identification Technology Research (CITeR) and the National Science Foundation under Grant No. 1822190.

\clearpage
\bibliographystyle{IEEEtran}
\bibliography{egbib}
\end{document}